\documentclass[10pt,letterpaper]{article}
\usepackage{cite}
\usepackage{opex3}
\usepackage{color}
\usepackage{braket}
\usepackage{microtype} 
\usepackage[hidelinks]{hyperref}
\usepackage{pgfplots}
\usepackage{gensymb}
\usepackage{amsmath}
\usepackage{amsfonts}
\usepackage{amssymb}
\usepackage{multicol}
\usepackage{comment} 
\usepackage{relsize}
\usepackage{xcolor}

\definecolor{colorbrewer1}{RGB}{55,126,184}
\definecolor{colorbrewer2}{RGB}{252,141,89}
\definecolor{OliveGreen}{cmyk}{0.64,0,0.95,0.40}

\begin{document}

\title{On the Resilience of Scalar and Vector Vortex Modes in Turbulence}
\author{Mitchell A. Cox$^{1,2}$, Carmelo Rosales-Guzm\'an$^{2}$, Martin P. J. Lavery$^{3}$,\\ Daniel J. Versfeld$^{1}$ and Andrew Forbes$^{2}$}
\address{$^1$School of Electrical and Information Engineering, University of the Witwatersrand, Johannesburg, 2050, South Africa\\
$^2$School of Physics, University of the Witwatersrand, Johannesburg, 2050, South Africa\\
$^3$School of Engineering, University of Glasgow, Lanarkshire, G12 8LT, UK}
\email{andrew.forbes@wits.ac.za}

\begin{abstract}
Free-space optical communication with spatial modes of light has become topical due to the possibility of dramatically increasing communication bandwidth via Mode Division Multiplexing (MDM). While both scalar and vector vortex modes have been used as transmission bases, it has been suggested that the latter is more robust in turbulence.  Using orbital angular momentum as an example, we demonstrate theoretically and experimentally that the crosstalk due to turbulence is the same in the scalar and vector basis sets of such modes. This work brings new insights about the behaviour of vector and scalar modes in turbulence, but more importantly it demonstrates that when considering optimal modes for MDM, the choice should not necessarily be based on their vectorial nature.
\end{abstract}


\bibliography{Ref_SvsV2}
\bibliographystyle{osajnl}

\section{Introduction}
\label{sec:intro}
Mode Division Multiplexing (MDM) was first suggested more than three decades ago as a technique to increase the bandwidth in fibre-based optical communications \cite{Berdague}. The realisation in recent years that current optical fibre communication systems will undergo a ``capacity crunch'' in the near future has renewed interest in the research \cite{Richardson1}.  Along this line, Orbital Angular Momentum (OAM) stands out as the mode of choice due to its topical nature and ease of measurement \cite{Willner,Forbes2016,Berkhout2010a,OSullivan2012}. Successful demonstrations in both optical fibres and free space \cite{Gibson2004,Krenn2014,Willner,Wang1,Funes2015,Bruning2016} suggest the viability of OAM-based optical communication systems, even though its capacity limits still remain controversial \cite{Andersson2015,Zhao}. 

A limiting factor to the deployment of free-space optical communications is turbulent atmospheric conditions which lead to aberrations of the optical wave-front of a propagating beam. These aberrations arise from localised changes in temperature and pressure, causing spatial variations in the refractive index of the atmosphere \cite{anguita2008,Pu2010,Chen2013b,Chen2016}. In MDM, turbulence causes degradation of the orthogonal spatial modes which results in the spreading of power into neighbouring modes. In the case of an MDM communication system, this so-called crosstalk imposes limits on the achievable channel capacity \cite{Rodenburg12,Milione2015,ren2013,Ren2016}. Investigation into the mitigation of the effects of atmospheric turbulence on beams carrying OAM have shown very promising results, where techniques employing specific modal selection, adaptive optics and digital signal processing have been demonstrated \cite{Zhao2012,Huang2014,Ren2014}. 


Vector vortex modes are non-separable states of light in which polarisation and OAM are coupled, which results in an inhomogeneous polarisation distribution \cite{milione2011higher}. Recently, the use of vector modes was proposed as an alternative to scalar modes to encode information \cite{Milione2015e}.  It has been shown numerically that non-uniformly polarised beams such as a vector vortex beams are analogous to partially coherent beams, in as far as their resilience to atmospheric turbulence is concerned \cite{Gu2009,Gu2012}. Another study has postulated that the polarisation distribution of a vector vortex beam is maintained even after its intensity distribution has degraded and thus a portion of the information encoded in polarisation is still present \cite{cheng2009}. From these studies it has been inferred that vector vortex beams are more resilient to turbulence as compared to their scalar counterparts \cite{Gu2009,Gu2012,cheng2009}.

Here we show that Cylindrical Vector Vortex (CVV) modes are not more resilient to atmospheric turbulence than their scalar (OAM) counterparts.  We confirm this experimentally by measuring the crosstalk between basis elements of both mode sets, perturbed by Kolmogorov, thin phase screens encoded on a polarisation invariant Spatial Light Modulator (SLM). In comparing the modal crosstalk induced in each case, it is determined that although the crosstalk between modes within the scalar vortex and CVV bases is distributed differently, the total crosstalk is in fact identical within experimental error. The coupling between OAM and polarization in CVV beams is not sufficient to make their phase variation less susceptible to atmospheric turbulence when compared to circularly polarized scalar vortex beams.

\section{Theory}
\label{sec:theory}

\begin{figure}[tb]
\centering
\includegraphics[width=0.75\textwidth]{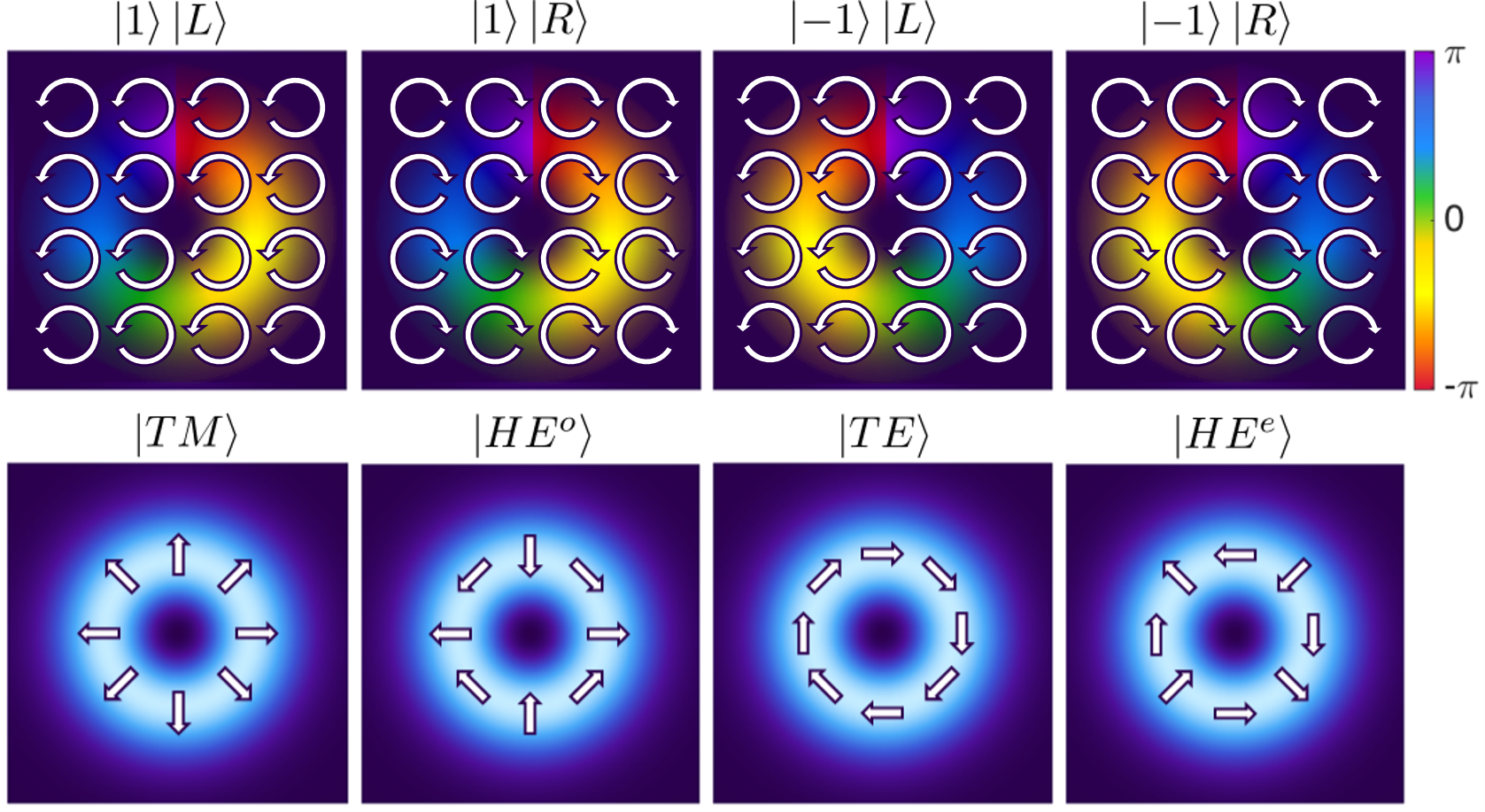}
\caption{\label{fig:modes}Illustration of the scalar (top) and CVV (bottom) modes described in Eq.~\ref{eq:scalar} and Eq.~\ref{eq:vector} respectively, with arrows indicating polarisation distribution which is constant for scalar and variable for CVV modes, for $\ell=\pm1$. The phase of the four scalar modes in the transverse plane increases from $-\pi$ (red) to $\pi$ (blue) for positive $\ell$'s and decreases in the opposite direction for negative $\ell$'s whereas the phase for CVV modes is a superposition of both.}
\end{figure}

A scalar basis set can be constructed by combining the degree of freedom provided by circular polarisation with the degrees of freedom given by the infinite set of OAM modes, illustrated in Fig.~\ref{fig:modes}. These modes can be written, using Dirac notation for brevity, as:

\begin{subequations}
\begin{multicols}{2}
\begin{equation}
\ket{R^+}=\ket{\ell}\ket{R},  
\end{equation}
\vspace*{-1.4\baselineskip}
\begin{equation}
\ket{L^+}=\ket{\ell}\ket{L},
\end{equation}
\begin{equation}
\ket{R^-}=\ket{-\ell}\ket{R},
\end{equation}
\vspace*{-1.4\baselineskip}
\begin{equation}
\ket{L^-}=\ket{-\ell}\ket{L},
\end{equation}  
\end{multicols}
\label{eq:scalar}
\end{subequations}

\noindent where, $R$ and $L$ denotes right and left circular polarization, respectively, and $\ell \in \mathbb{Z}$ relates to the amount of orbital angular momentum,  $\ell\hbar$, per photon\cite{Allen92}. The elements of this set are orthogonal to each other. 

CVV modes are non-separable states of light where polarization and OAM modes are coupled. A basis set of four orthogonal CVV modes can be constructed as linear combinations of the scalar mode set in Eq.~\ref{eq:scalar}. The resulting modes carry no overall angular momentum and their transverse polarization state is not constant, illustrated in the bottom row of Fig.~\ref{fig:modes}. Akin to the scalar case, these modes can be represented in the circular polarization basis using Dirac notation as:

\vspace*{-2\baselineskip}
\begin{subequations}
\begin{multicols}{2}
\begin{equation}
\ket{TM} = \frac{1}{\sqrt{2}}(\ket{\ell}\ket{R} + \ket{-\ell}\ket{L}),
\end{equation}
\vspace*{-1.8\baselineskip}
\begin{equation}
\ket{TE} = \frac{1}{\sqrt{2}}(\ket{\ell}\ket{R} - \ket{-\ell}\ket{L}),
\end{equation}
\begin{equation}
\ket{HE^e} = \frac{1}{\sqrt{2}}(\ket{\ell}\ket{L} + \ket{-\ell}\ket{R}),
\end{equation}
\vspace*{-1.8\baselineskip}
\begin{equation}
\ket{HE^o} = \frac{1}{\sqrt{2}}(\ket{\ell}\ket{L} - \ket{-\ell}\ket{R}).
\end{equation}
\end{multicols}
\label{eq:vector}
\end{subequations}

Each mode in a given set (vector and scalar) is orthogonal with the others in the set, and mutually unbiased across the sets.  Moreover, each of the modes described by Eq.~\ref{eq:scalar} and \ref{eq:vector} can be represented on a high order Poincar\'e sphere \cite{milione2011higher,Naidoo2016}.

\subsection{Inter-Mode Crosstalk}
\label{subsec:crosstalk}

In the presence of atmospheric turbulence, the polarisation of a beam is not affected because the atmosphere is not birefringent, which can readily be validated by tilting ones head (e.g., Kolmogorov turbulence assumes a homogeneous and isotropic atmosphere). The spatial degree of freedom, however, experiences aberrations which result in the coupling of modes into neighbouring modes, degrading their orthogonality. The amount of mode coupling or crosstalk is dependent on the strength of the turbulence in the channel, which can be expressed, in general, by:

\begin{equation}
\label{eq:turbSum}
\ket{\ell} \xrightarrow{turb.} \sum_{\ell'}{p_{\ell-\ell'}\ket{\ell'}},
\end{equation}

\noindent where $p_{\ell-\ell'}$ are the mode coupling weightings described by some distribution (e.g., $p_0$ would represent the modal power in the original OAM mode). This is used to find general expressions for what a given mode propagating through turbulence will transform into. For instance, the final state of the scalar mode $\ket{R^+}$ will be given by:

\begin{equation}
\label{eq:scalarTurb}
\ket{R^+} \xrightarrow{turb.} p_0 \ket{\ell} \ket{R} + p_{2\ell} \ket{-\ell}\ket{R} = \ket{R^+_{turb.}}
\end{equation}

\noindent Analogous expressions can be found for $\ket{R^-}$, $\ket{L^+}$  and $\ket{L^-}$. Similarly, applying Equation~\ref{eq:turbSum} to a vector mode $\ket{TM}$:

\begin{equation}
\label{eq:vectorTurb}
\ket{TM} \xrightarrow{turb.} p_0 \ket{\ell} \ket{R} + p_0 \ket{-\ell}\ket{L} + p_{-2\ell} \ket{\ell} \ket{L} +  p_{2\ell} \ket{-\ell}\ket{R} = \ket{TM_{turb}}.
\end{equation}

\noindent Analogous relations for $\ket{TE}$, $\ket{HE^e}$ and $\ket{HE^o}$ can also be found. 

The crosstalk for each input mode is the magnitude of the inner product between each input mode and all the expected output modes after turbulence. By way of example, the inner product of the scalar mode $\ket{R^+}$ with the mode $\ket{R^+_{turb}}$, will be:

\begin{equation}
||\braket{R^+|R^+_{turb}}||^2 =||\bra{\ell}\bra{R}(p_0 \ket{\ell} \ket{R} + p_{2\ell} \ket{-\ell}\ket{R})||^2= ||P_0||^2.
\end{equation}

\noindent Similarly, the inner product of the vector mode $\ket{TM}$ with the mode $\ket{TM_{turb}}$ will be:

\begin{equation}
\begin{aligned}
||\braket{TM|TM_{turb}}||^2 &=||(\frac{1}{\sqrt{2}}(\bra{\ell}\bra{R} + \bra{-\ell}\bra{L})(p_0 \ket{\ell} \ket{R} p_0 \ket{-\ell}\ket{L} + p_{-2\ell} \ket{\ell} \ket{L}\\
& +p_{2\ell} \ket{-\ell}\ket{R})||^2=||P_{0}||^2
\end{aligned}
\end{equation}

\noindent The crosstalk terms for all of the modes can be summarised in the form of a matrix as:

\begin{equation}
\label{eq:scalarCrosstalk}
M_{scalar}=
  \begin{pmatrix}
		 ||p_0||^2 & 0 & ||p_{-2\ell}||^2 & 0 \\ 
		 0 & ||p_0||^2 & 0 & ||p_{-2\ell}||^2 \\ 
		 ||p_{2\ell}||^2 &0  & ||p_0||^2 & 0 \\ 
		 0 & ||p_{2\ell}||^2 & 0 & ||p_0||^2 \\
  \end{pmatrix},
\end{equation}

\begin{equation}
\label{eq:vectorCrosstalk}
M_{vector}=
  \begin{pmatrix}		 
		 ||p_0||^2 & \frac{||p_{-2\ell}+p_{2\ell}||^2}{4}& 0 & \frac{||p_{-2\ell}-p_{2\ell}||^2}{4} \\ 
         
         \frac{||p_{-2\ell}+p_{2\ell}||^2}{4}& ||p_0||^2 &\frac{||(p_{-2\ell}-p_{2\ell}||^2}{4}   & 0 \\
         
		0 & \frac{||p_{-2\ell}-P_{2\ell}||^2}{4} & ||p_0||^2 & \frac{||p_{-2\ell}+P_{2\ell}||^2}{4} \\ 
        
		\frac{||p_{-2\ell}-p_{2\ell}||^2}{4} & 0 & \frac{||p_{-2\ell}+p_{2\ell}||^2}{4} & ||p_0||^2 \\
        
  \end{pmatrix}.
\end{equation}

\noindent Notice that the terms in the diagonal are identical both cases. Under zero turbulence, where the mode coupling weightings for the crosstalk terms can be assumed to be zero ($p_{2\ell}=p_{-2\ell}=0$), only the terms in the diagonal remain. This is consistent with the notion that when propagating in free-space with no aberrations induced by atmospheric turbulence, the output mode equals the input mode.  

The total crosstalk, $N$, in each case can be computed by adding the off-diagonal elements as: 

\begin{equation}
N_{vector}=\sum_{i\neq j} M_{vector}=(||p_{2\ell}+p_{-2\ell}||^2+||p_{2\ell}-p_{-2\ell}||^2) = 2(||p_{-2\ell}||^2+||p_{2\ell}||^2)
\end{equation}

\begin{equation}
N_{scalar}=\sum_{i\neq j} M_{scalar}=2(||p_{-2}||^2+||p_{2}||^2)
\end{equation}

\noindent which therefore leads to our claim that the total crosstalk in each case is in fact identical:
    
\begin{equation}
N_{scalar}=N_{vector}= 2(||p_{-2}||^2+||p_{2}||^2)
\end{equation}

\section{Experiment} 
\label{sec:experiment}

\begin{figure}
\centering
\includegraphics[width=\textwidth]{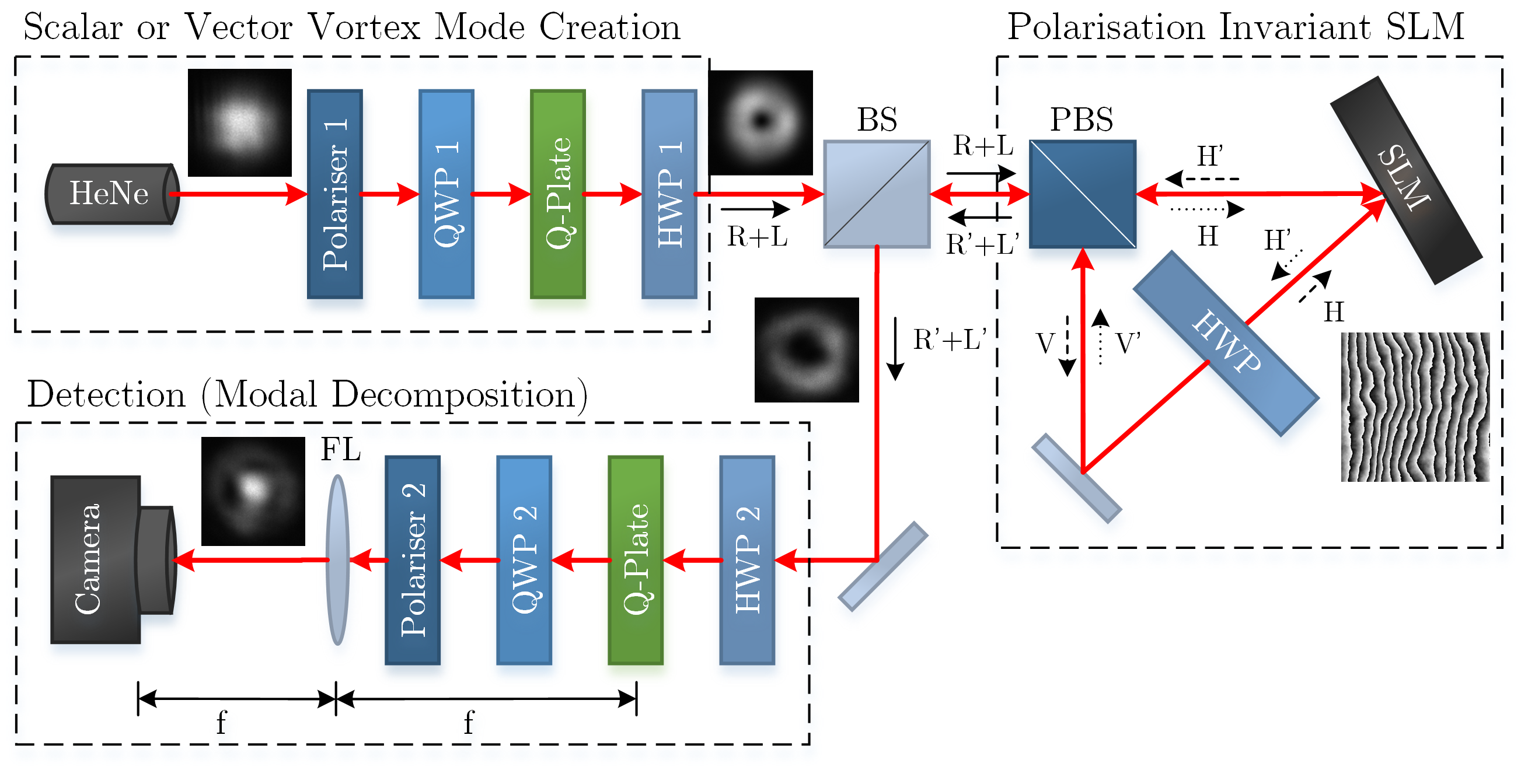}
\caption{\label{fig:diagram}Simplified schematic diagram of the experimental setup showing the three main parts. The configuration of the half- and quarter-wave plates (HWP and QWP) and polarisers for different modes is summarised in the Appendix. The polarisation invariant SLM is made up by a polarising beam splitter (PBS), mirror and half wave plate rotated to $45\degree$ so that arbitrarily polarised scalar and vector beams can be modulated by the SLM, which is encoded with random Kolmogorov turbulence screens. The perturbed beams which return along the same path as the incoming beams are directed by a beam splitter (BS) to the detection part of the setup which performs modal decomposition.}
\end{figure}

Figure~\ref{fig:diagram} shows the experimental setup used to corroborate our theoretical findings. It consisted of three main stages: generation of scalar and CVV modes, turbulence using a Spatial Light Modulator (SLM) and finally, detection. In the first stage, both scalar and CVV modes were generated using a $q$-plate ($q=1/2$) in conjunction with polarisers and wave plates \cite{Marrucci2006}. The configuration of each element is detailed in the Appendix. For the second stage, a HoloEye Pluto SLM was used (PLUTO-VIS, $1920\times1080$ pixels with $8~\mu\mathrm{m}$ pitch, calibrated for a $2\pi$ maximum phase shift at 633~nm) to simulate atmospheric turbulence. Random Kolmogorov phase screens were used, specified by their Strehl Ratio (SR) \cite{Lane1992}. Since this SLM is only able to modulate horizontally polarised light, a polarisation invariant arrangement was implemented, as illustrated in Fig.~\ref{fig:diagram}, following the approach in ref. \cite{Liu2015}. In the last stage, the perturbed modes were detected by inverting the creation stage (exploiting the reciprocity of light) followed by an inner product measurement to quantitatively infer the mode coupling weightings by using a technique known as modal decomposition \cite{Forbes2016,Kaiser2009}. 

Each input mode was perturbed by one hundred discrete instances of thin phase atmospheric turbulence of a specific strength. The strength of turbulence was increased in increments of $0.1$ from $SR=0.1$ to $1.0$. The channel matrices, $M_{scalar}$ and $M_{vector}$, described in Section~\ref{sec:theory} were then generated for each turbulence strength using the averaged and normalised intensity data for comparison to the theoretical calculations in Eq.~\ref{eq:scalarCrosstalk} and \ref{eq:vectorCrosstalk}.

\section{Results and Discussion}
\label{sec:results} 

\begin{figure}
\centering
\begin{tikzpicture}
    \begin{axis}[
    	title={Scalar},
        width=3cm,
        height=3cm,
        scale only axis,
        enlargelimits=false,
        axis on top,
        ticklabel style = {font=\small},
        xtick={10,30,50,70},
        ytick={10,30,50,70},
        ytick style={draw=none},
        xtick style={draw=none},
        xlabel={Input Mode},
        ylabel={Output Mode},
    	xticklabels={$R^-$,$L^-$,$R^+$,$L^+$},
        yticklabels={$L^+$,$R^+$,$L^-$,$R^-$},
    ]
      \addplot graphics[xmin=0,xmax=80,ymin=0,ymax=80] {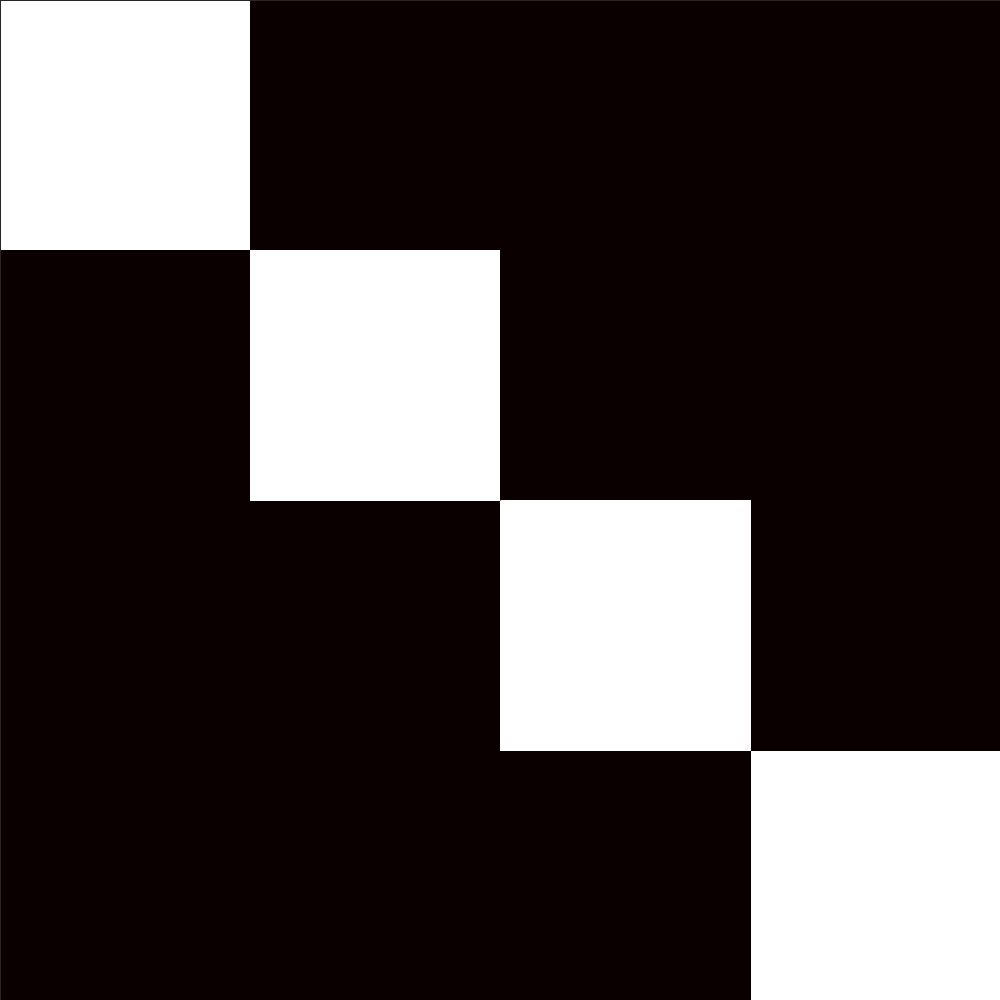};
    \end{axis}
  \end{tikzpicture}%
  \qquad%
  \begin{tikzpicture}
    \begin{axis}[
        title={Vector},
        colormap/hot2, 
		colorbar,
        colorbar style={ytick={0.0,0.2,0.4,0.6,0.8,1.0},yticklabel style={
            /pgf/number format/.cd,
                fixed,
                fixed zerofill, ,precision=1
        }},
        width=3cm,
        height=3cm,
        scale only axis,
        enlargelimits=false,
        axis on top,
        ticklabel style = {font=\small},
        xtick={10,30,50,70},
        ytick={10,30,50,70},
        ytick style={draw=none},
        xtick style={draw=none},
        xlabel={Input Mode},
    	xticklabels={$\ket{TM}$,$\ket{HE^e}$,$\ket{TE}$,$\ket{HE^o}$},
        yticklabels={$\ket{HE^o}$,$\ket{TE}$,$\ket{HE^e}$,$\ket{TM}$},
    ]
      \addplot graphics[xmin=0,xmax=80,ymin=0,ymax=80] {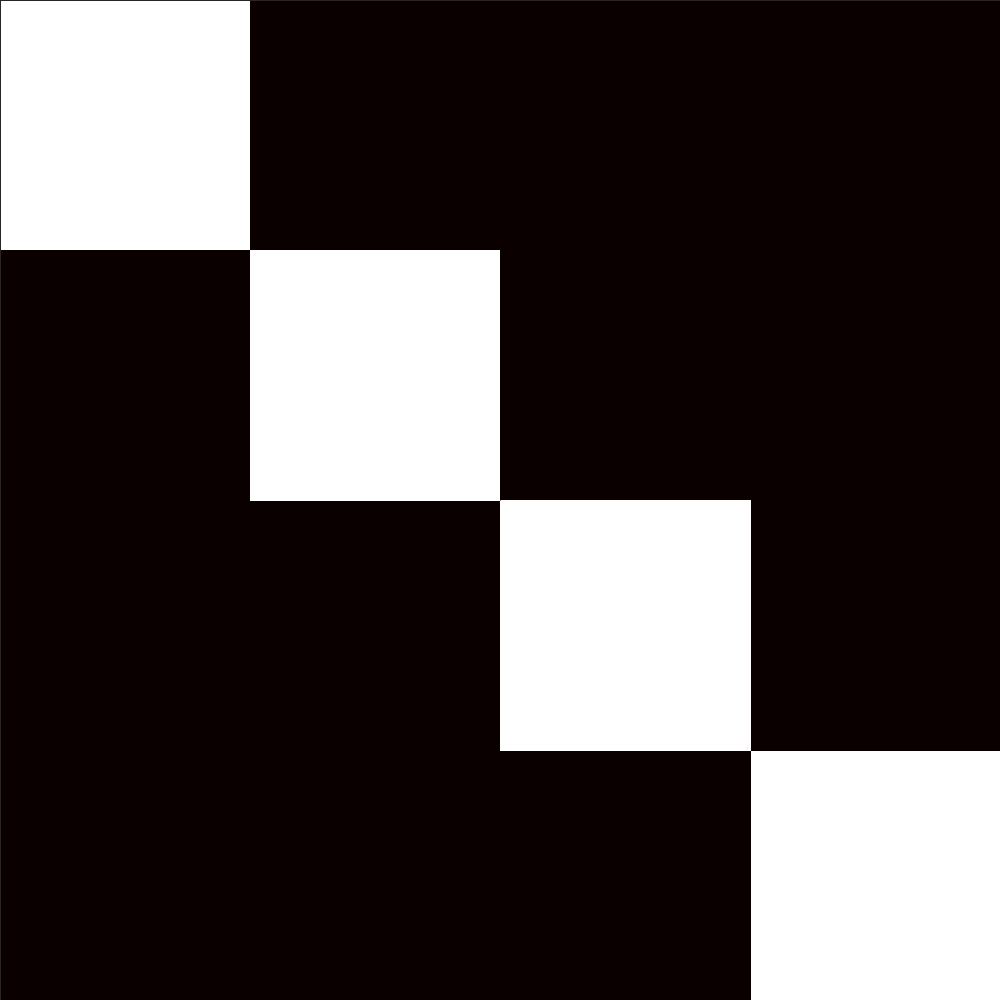};
    \end{axis}
  \end{tikzpicture}  
  \caption{\label{fig:crosstalkBaseline}Vector and scalar measurements showing no crosstalk with no turbulence ($SR=1.0$) for verification of the experimental setup.}
\end{figure}

The experimental setup was verified by encoding the SLM with $SR=1.0$ turbulence, which is simply a grating. This is the control measurement for the experiment and the crosstalk was zero, as expected, shown in Figure~\ref{fig:crosstalkBaseline}. 

\begin{figure}
\centering
\begin{tikzpicture}
    \begin{axis}[
    	title={Scalar},
        width=3cm,
        height=3cm,
        scale only axis,
        enlargelimits=false,
        axis on top,
        ticklabel style = {font=\small},
        xtick={10,30,50,70},
        ytick={10,30,50,70},
        ylabel={Output Mode},
        ytick style={draw=none},
        xtick style={draw=none},
    	xticklabels={$R^-$,$L^-$,$R^+$,$L^+$},
        yticklabels={$L^+$,$R^+$,$L^-$,$R^-$},
    ]
      \addplot graphics[xmin=0,xmax=80,ymin=0,ymax=80] {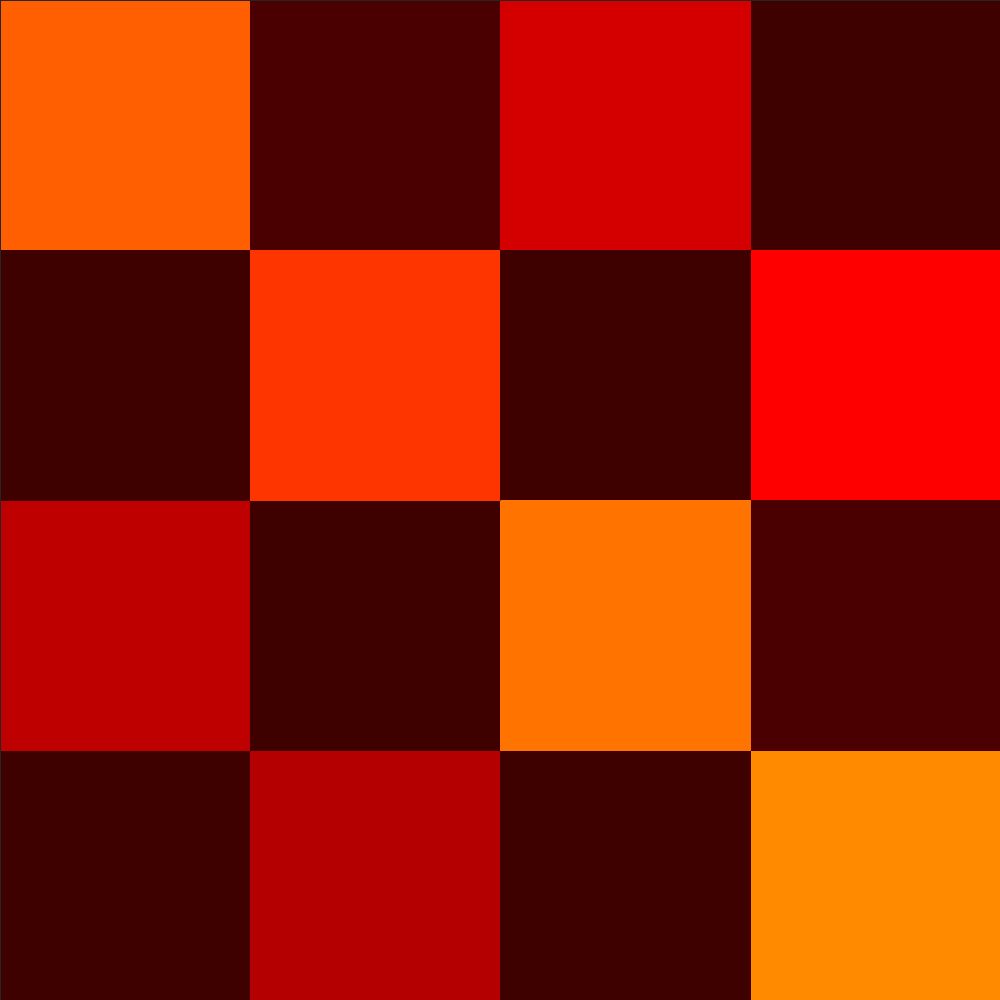};
    \end{axis}
  \end{tikzpicture}%
  \qquad%
  \begin{tikzpicture}
    \begin{axis}[
        title={Vector},
        colormap/hot2, 
		colorbar,
        colorbar style={ytick={0.0,0.2,0.4,0.6,0.8,1.0},yticklabel style={
            /pgf/number format/.cd,
                fixed,
                fixed zerofill, ,precision=1
        }},
        width=3cm,
        height=3cm,
        scale only axis,
        enlargelimits=false,
        axis on top,
        ticklabel style = {font=\small},
        xtick={10,30,50,70},
        ytick={10,30,50,70},
        ytick style={draw=none},
        xtick style={draw=none},
    	xticklabels={$\ket{TM}$,$\ket{HE^e}$,$\ket{TE}$,$\ket{HE^o}$},
        yticklabels={$\ket{HE^o}$,$\ket{TE}$,$\ket{HE^e}$,$\ket{TM}$},
    ]
      \addplot graphics[xmin=0,xmax=80,ymin=0,ymax=80] {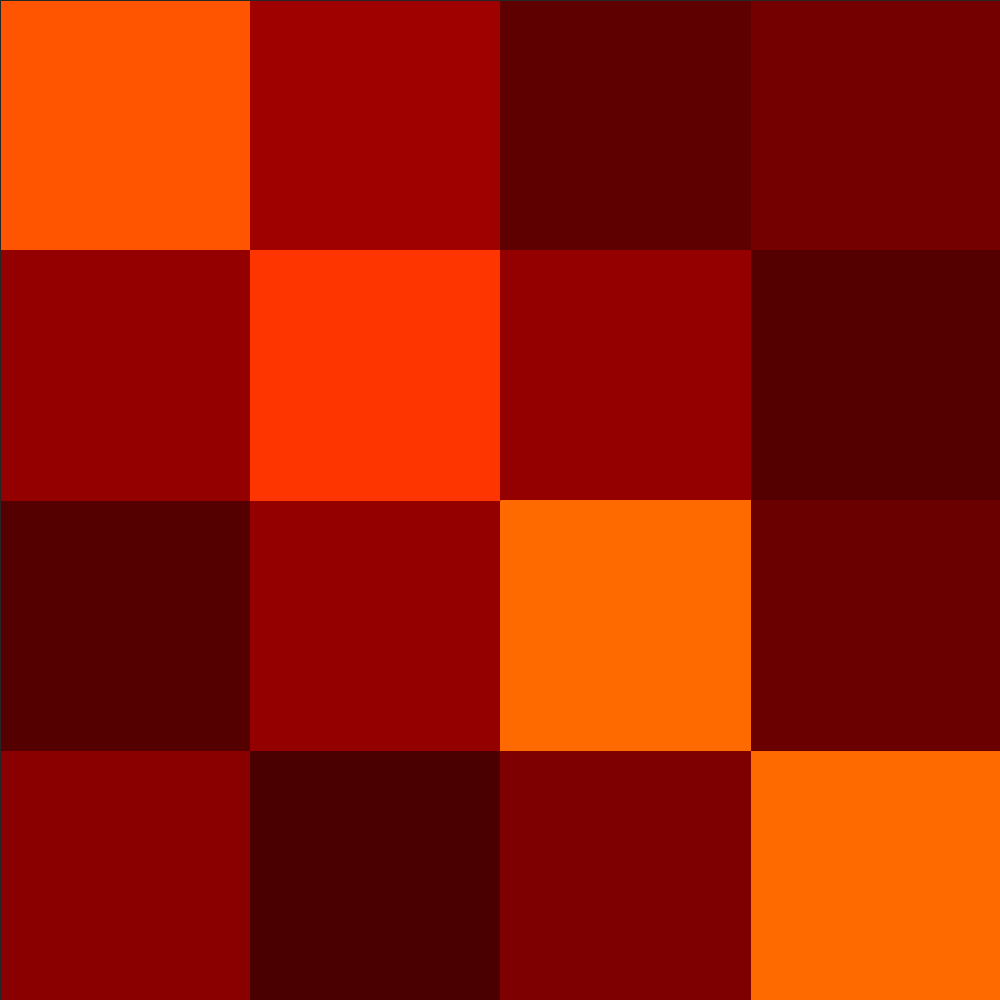};
    \end{axis}
  \end{tikzpicture}
  
  \begin{tikzpicture}
    \begin{axis}[
        width=3cm,
        height=3cm,
        scale only axis,
        enlargelimits=false,
        axis on top,
        ticklabel style = {font=\small},
        xtick={10,30,50,70},
        ytick={10,30,50,70},
        xlabel={Input Mode},
        ylabel={Output Mode},
        ytick style={draw=none},
        xtick style={draw=none},
    	xticklabels={$R^-$,$L^-$,$R^+$,$L^+$},
        yticklabels={$L^+$,$R^+$,$L^-$,$R^-$},
    ]
      \addplot graphics[xmin=0,xmax=80,ymin=0,ymax=80] {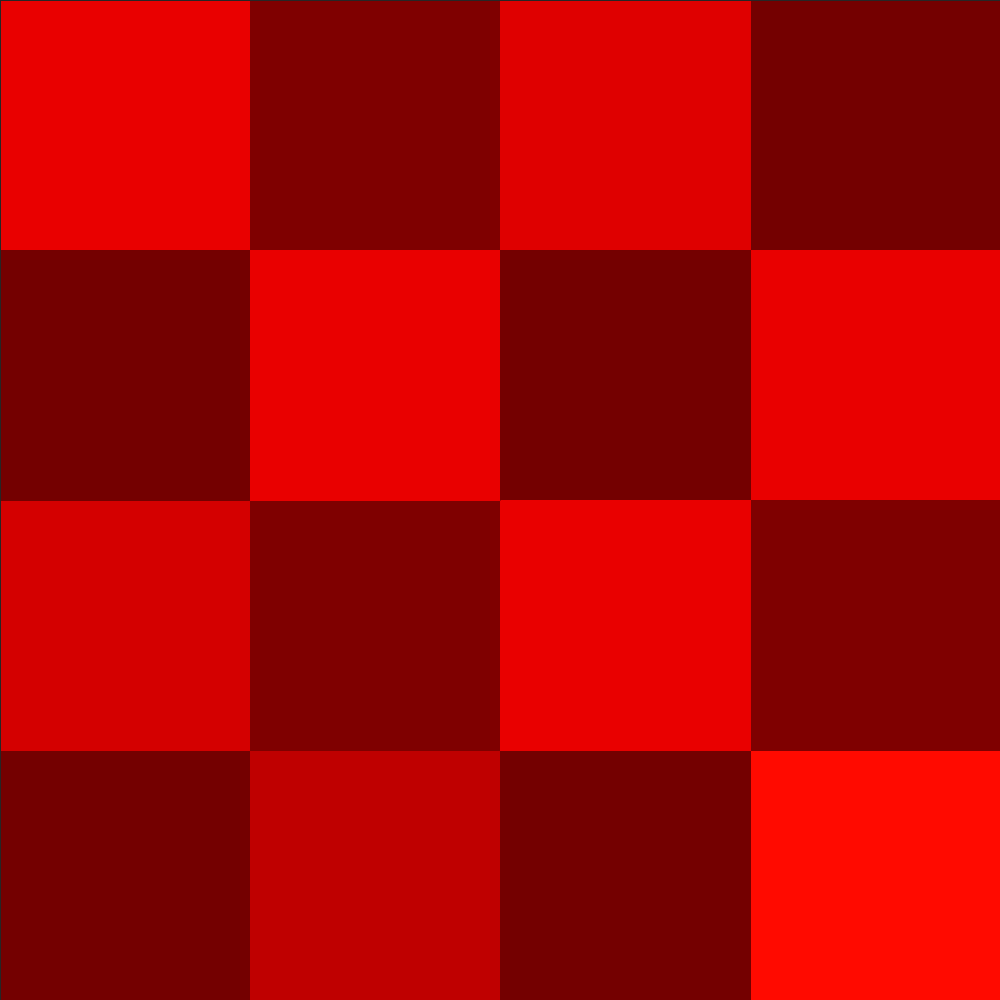};
    \end{axis}
  \end{tikzpicture}%
  \qquad%
  \begin{tikzpicture}
    \begin{axis}[
    	colormap/hot2, 
		colorbar,
        colorbar style={ytick={0.0,0.2,0.4,0.6,0.8,1.0},yticklabel style={
            /pgf/number format/.cd,
                fixed,
                fixed zerofill, ,precision=1
        }},
        width=3cm,
        height=3cm,
        scale only axis,
        enlargelimits=false,
        axis on top,
        ticklabel style = {font=\small},
        xtick={10,30,50,70},
        ytick={10,30,50,70},
        xlabel={Input Mode},
        ytick style={draw=none},
        xtick style={draw=none},
    	xticklabels={$\ket{TM}$,$\ket{HE^e}$,$\ket{TE}$,$\ket{HE^o}$},
        yticklabels={$\ket{HE^o}$,$\ket{TE}$,$\ket{HE^e}$,$\ket{TM}$},
    ]
      \addplot graphics[xmin=0,xmax=80,ymin=0,ymax=80] {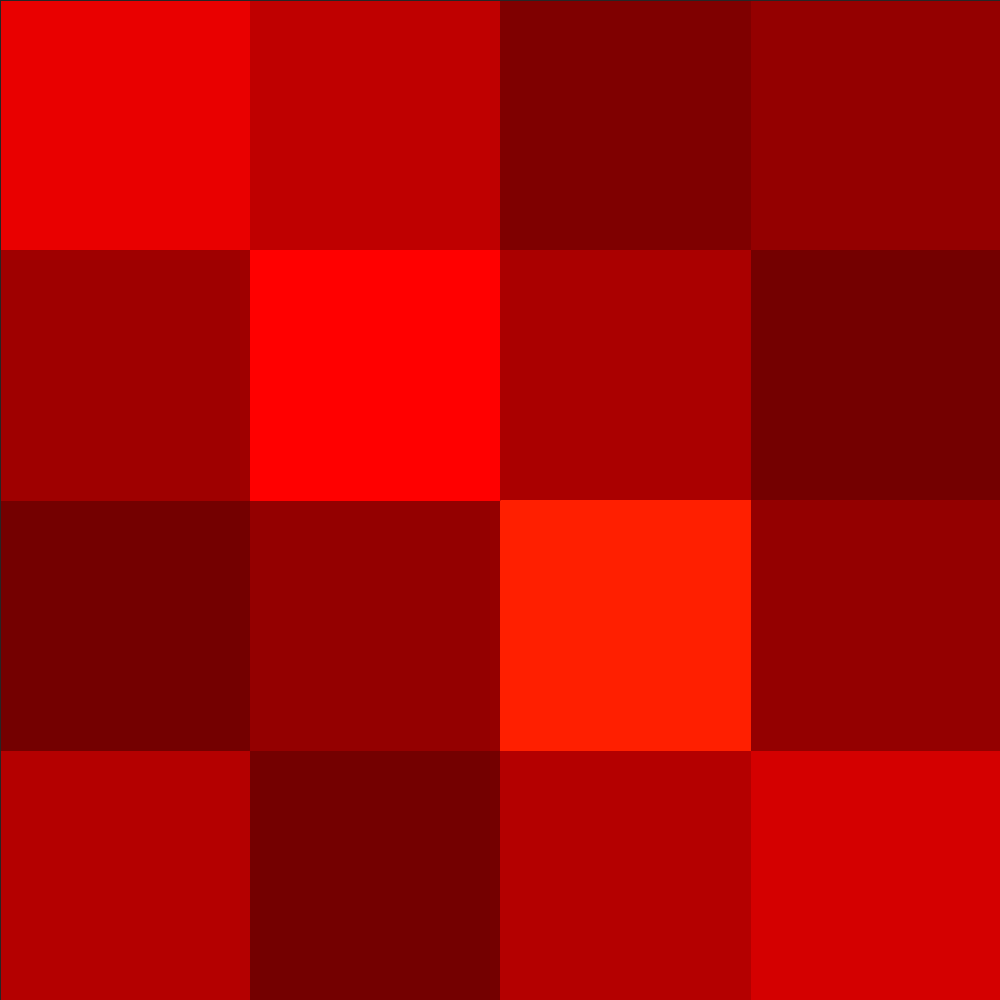};
    \end{axis}
  \end{tikzpicture}
  \caption{\label{fig:crosstalk2x2}Comparison of vector and scalar crosstalk for $SR=0.6$ (above) and $SR=0.2$ (below).}
  \end{figure}

Figure~\ref{fig:crosstalk2x2} shows the channel matrix comparison for $SR=0.6$ and 0.2 of scalar and vector cases. These two Strehl Ratio's are arbitrary but were chosen because they demonstrate cases for medium and strong turbulence and the clear effect of stronger turbulence, where power is spread from the signal on the diagonal into the crosstalk elements. 

According to the theory and the assumption that crosstalk does not occur across polarisation states, some elements in the matrices shown in Fig.~\ref{fig:crosstalk2x2} should be black when in fact they show a small amount of signal. This is attributed to the normalisation of the low dynamic range signal in stronger turbulence and the fact that the PBS is not 100\% efficient.

\begin{figure}
	\centering
	\begin{tikzpicture}
    \node[anchor=south west,inner sep=0pt] at (-0.03,3.42)
    {\includegraphics[width=11.745cm]{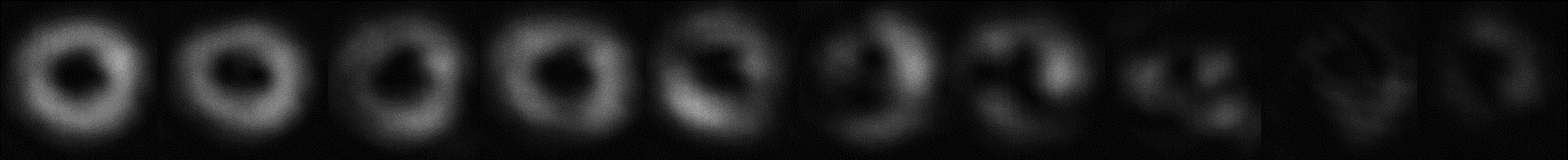}};
	\begin{axis}[ %
	grid=major,
	grid style={dashed,gray!20},
	width={1\textwidth},
	height=5cm,
	legend entries={Scalar, Vector},
	legend cell align=left,
	legend style={draw=none},
	legend pos = north west,
	xlabel={Strehl Ratio at SLM},
    xticklabel style={/pgf/number format/.cd,fixed,fixed zerofill,precision=1},
    scaled x ticks=false,
	minor x tick num=1,
	minor y tick num=1,
	xmin=0.05,
	xmax=1.05,
	ymin=0,
	ymax=80,
	ylabel={Crosstalk (\%)},
    x dir=reverse,
    axis line style=thick
	] %
	
    \addplot[very thick,colorbrewer1,mark=*,mark options={fill=white},/pgfplots/error bars/.cd,x dir=none,y dir=both,y explicit,error mark options={
      rotate=90,
      mark size=2pt,
      line width=1pt
    }]
	table[x index=0, y expr=\thisrowno{5}*100, y error expr=\thisrowno{10}*100, col sep=comma] {scalar.csv}; 
    
    \addplot[very thick,colorbrewer2,mark=square*, mark size=1.75,mark options={fill=white},/pgfplots/error bars/.cd,x dir=none,y dir=both,y explicit,error mark options={
      rotate=90,
      mark size=2pt,
      line width=1pt
    }]
	table[x index=0, y expr=\thisrowno{5}*100, y error expr=\thisrowno{10}*100, col sep=comma] {vector.csv};
	
	\end{axis} %
	\end{tikzpicture}
	
	\caption{\label{fig:NoiseSummary} Percentage of the signal in the crosstalk of scalar and vector cases showing insets of the beam with increasing turbulence. Both mode sets have the same crosstalk performance. Error bars are the standard error of the mean.}
	
\end{figure}

In order to compare the scalar and vector cases against each other, the crosstalk elements for each input mode were added to each other according to the theory in Section~\ref{subsec:crosstalk}. The sum is then normalised, resulting in a total average crosstalk percentage for each turbulence strength for each mode set, visible in Figure~\ref{fig:NoiseSummary}.

The crosstalk performance for both scalar and vector modes is identical within the measurement error for all turbulence strengths. These results agree well with the theory presented in Section~\ref{sec:theory} and indicate that there is no performance benefit to using CVV modes over scalar vortex modes in a thin phase turbulence regime with no birefringence. It should be noted that a similar approach may be used for other basis sets of vector and scalar modes, however, the results in this paper cannot be extended to modes with asymmetric spatial indices such as $\ket{\ell}\ket{R} + \ket{p}\ket{L}$, for example, as their resilience to turbulence cannot be assumed to be equal. 

A useful consequence of this similarity is that techniques that are use to mitigate turbulence effects on scalar beams may also be implemented for CVV beams. If a vector vortex mode is used in an MDM communication system, the additional degree of freedom afforded by polarisation is occupied and cannot be used for Polarisation Division Multiplexing (PDM). Therefore, as we have shown that the crosstalk performance between CVV and scalar vortex modes in thin phase atmospheric turbulence is identical, we suggest that the use of CVV modes may in fact limit the possible information capacity of a communication system.

\section{Conclusion}
\label{sec:conclusion}

It has been inferred in the literature that vector vortex modes are more resilient to atmospheric turbulence than their scalar counterparts. Here we define two similar scalar and cylindrical vector vortex mode bases and theoretically show that their crosstalk in turbulence is the same. This result was then verified experimentally in the thin phase, Kolmogorov regime with various turbulence strengths. The experimental results show identical crosstalk performance within the experimental error.

\section{Acknowledgements}
We would like to thank Bienvenu Ndagano for many fruitful conversations as well as Lorenzo Marrucci for providing the $q$-plates. The financial assistance of the National Research Foundation (NRF), the Claude Leon Foundation and CONACyT towards this research is acknowledged.

\newpage
\appendix
\section*{Appendix}
\label{sec:appendix}
\section{Generation of scalar and CVV modes}

The generation of both scalar and CVV modes was accomplished using a $q$-plate ($q=1/2$) in combination with quarter- and half- wave plates according to the unitary operation  $\hat{Q}$ of a $q$-plate:

\begin{subequations}
\label{eq:qTransform}
\begin{multicols}{2}
\begin{equation}
\hat{Q}
\begin{bmatrix}
    \ket{R}\\
    \ket{L} 
  \end{bmatrix}
  =
  \begin{bmatrix}
    \ket{+2q}\ket{L}\\
    \ket{-2q}\ket{R} 
  \end{bmatrix},
\end{equation}
\begin{equation}
\hat{Q}
\begin{bmatrix}
    \ket{H}\\
    \ket{V} 
  \end{bmatrix}
  =
  \begin{bmatrix}
    \ket{+2q}\ket{R} + \ket{-2q}\ket{L}\\
    \ket{+2q}\ket{R} - \ket{-2q}\ket{L}
  \end{bmatrix},
\end{equation}  
\end{multicols}
\end{subequations}

\noindent where $q$ is the charge of the topological singularity of the $q$-plate, related to the OAM by $\ell=2q$. When the input beam is circularly polarized, the output is a scalar beam of the opposite polarization embedded with an OAM of $\pm\ell$. Alternatively, when the input is linearly polarised (horizontal or vertical) the output is a CVV, in which, polarization and OAM are coupled. Experimentally, both sets of modes were generated using a q-plate, linear polarisers, half- and quarter- wave plates orientated at angles as summarized in Tab.~\ref{tab:exptConfig}. The detection of the modes was achieve by reversing the generation process using a similar configuration an similar optical elements, the orientation angles of the same are also summarised in this table.

\begin{table}
\centering
	\caption{\label{tab:exptConfig}Configuration of the various components of the experimental setup, depending on the required mode to be generated or decomposed.}
	\begin{tabular}{rcccc}
    	\hline
		Mode & Polariser 1 \& 2 & HWP 1 \& 2 & QWP 1 \& 2 \\ 
        \hline
		$\ket{1}\ket{R}$ & H & $0\degree$ & $45\degree$ \\ 
		$\ket{1}\ket{L}$ & H & - & $45\degree$ \\ 
		$\ket{-1}\ket{R}$ & V & - & $45\degree$ \\ 
		$\ket{-1}\ket{L}$ & H & $0\degree$ & $45\degree$ \\ 
		$\ket{TM}$ & H & - & - \\ 
		$\ket{TE}$  & V & - & - \\ 
		$\ket{HE^e}$  & H & $0\degree$ & -  \\ 
		$\ket{HE^o}$  & V & $0\degree$ & -  \\ 
        \hline
	\end{tabular} 
\end{table}

\end{document}